\documentclass[aps,prb,reprint,superscriptaddress]{revtex4-1}
\usepackage[dvips]{graphicx}
\usepackage{amsmath}
\usepackage{color}

\begin{document}

\title{Second harmonic microscopy of monolayer MoS$_2$}

\author{Nardeep Kumar}
\affiliation{Department of Physics and Astronomy, The University of Kansas, Lawrence, Kansas 66045, USA}

\author{Sina Najmaei}
\affiliation{Department of Mechanical Engineering and Materials Science, Rice University, Houston, Texas 77005, USA}

\author{Qiannan Cui}
\affiliation{Department of Physics and Astronomy, The University of Kansas, Lawrence, Kansas 66045, USA}

\author{Frank Ceballos}
\affiliation{Department of Physics and Astronomy, The University of Kansas, Lawrence, Kansas 66045, USA}

\author{Pulickel M. Ajayan}
\affiliation{Department of Mechanical Engineering and Materials Science, Rice University, Houston, Texas 77005, USA}

\author{Jun Lou}
\affiliation{Department of Mechanical Engineering and Materials Science, Rice University, Houston, Texas 77005, USA}

\author{Hui Zhao}
\email{huizhao@ku.edu}
\affiliation{Department of Physics and Astronomy, The University of Kansas, Lawrence, Kansas 66045, USA}

\date{\today}

\begin{abstract}
We show that the lack of inversion symmetry in monolayer MoS$_2$ allows strong optical second harmonic generation. Second harmonic of an 810-nm pulse is generated in a mechanically exfoliated monolayer, with a nonlinear susceptibility on the order of 10$^{-7}$ m/V. The susceptibility reduces by a factor of seven in trilayers, and by about two orders of magnitude in even layers.  A proof-of-principle second harmonic microscopy measurement is performed on samples grown by chemical vapor deposition, which illustrates potential applications of this effect in fast and non-invasive detection of crystalline orientation, thickness uniformity, layer stacking, and single-crystal domain size of atomically thin films of MoS$_2$ and similar materials.

\end{abstract}

\maketitle

Recently, there is a growing interest in exploring new types of atomically thin crystals based on layered materials, such as transition metal dichalcogenides.\cite{nn7699} The most extensively studied member of this family is MoS$_2$. In 2010, photoluminescence experiments\cite{l105136805,nl101271} and microscopic calculations\cite{b83245213,nl101271} indicated that, although bulk MoS$_2$ is an indirect semiconductors, its monolayer is a direct semiconductor with a bandgap of about 1.88 eV. Such a large bandgap and the structural similarity with widely studied graphene immediately stimulated investigations on its potential applications in logic electronics.\cite{ieeeted583042,nl113768} In 2011, top-gated transistors based on MoS$_2$ monolayers were fabricated.\cite{nn6147} Later, integrated circuits based on monolayer\cite{acsnano59934} and bilayer\cite{nl124674} for logic operations were demonstrated. Ambipolar transport in a multilayer transistor gated by ionic liquids was also demonstrated, showing the feasibility to develop p-n-junction-based devices.\cite{nl121136} Furthermore, the recently demonstrated superior strength and flexibility\cite{acsnano59703} make MoS$_2$ atomic layers an attractive candidate for applications in flexible electronics.\cite{small82994,nl124013} In addition, since monolayer MoS$_2$ has a bandgap in the visible range, has workfunctions that are compatible with commonly used electrode materials, and has stable charge exciton state even at room temperature,\cite{nm12207} it is also an attractive candidate for various optoelectronic and photonic applications, such as phototransistors\cite{acsnano674,nl123695} and heterojunction solar cells.\cite{apl100153901} 

In contrast to these breakthroughs in understanding the mechanical, electronic, and linear optical properties of MoS$_2$ atomic layers, little is known about their {\it nonlinear} optical properties. Nonlinear optical responses are important aspects of light-matter interaction, and can play important roles in various photonic and optoelectronic applications, especially in those involving high intensity laser beams. Bulk MoS$_2$ crystal with $2H$ stacking order belongs to space group $D_{6h}$, which is inversion symmetric. Hence, its second-order nonlinear response should vanish.\cite{bookboyd} Indeed, one early experiment showed that second-order nonlinear susceptibility of $2H$ bulk MoS$_2$ is at most 10$^{-14}$ m/V.\cite{josab151017} However, the inversion symmetry is broken in a monolayer, which has $D_{3h}$ symmetry. One consequence of such a symmetry reduction is to allow valley-selective optical interband transitions, which has been observed by several groups recently\cite{nc3887,nnano7490,nnano7494,l108196802,b86081301,nm12207} and can be used for valleytronics, in which the valley index of electrons is used to carry information.  Here we show that the lack of inversion symmetry allows unusually strong optical second harmonic generation (SHG) in monolayer MoS$_2$ flakes prepared by mechanical exfoliation and chemical vapor deposition (CVD). This effect is very sensitive to layer thickness, crystalline orientation, and layer stacking. Based on these properties, we demonstrate a second harmonic microscopy for characterization of MoS$_2$ thin films.

\begin{figure}
\centering
\includegraphics[width=8.5cm]{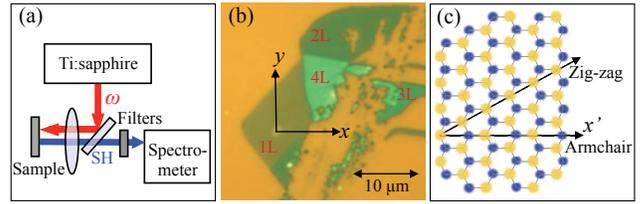}
\caption{(a) Schematics of the experimental setup. (b) Microscope image of a mechanically exfoliated MoS$_2$ flake. (c) Lattice structure of monolayer MoS$_2$. }
\label{fig:setup}
\end{figure}

Figure \ref{fig:setup}(a) shows the experimental setup. The fundamental pulse with an angular frequency $\omega$ and a central wavelength of 810 nm is obtained from a Ti:sapphire laser. It is tightly focused to a spot of 2~$\mu$m (full width at half maximum) by a microscope objective lens. The second harmonic (SH) generated is collected by the same lens, and detected by a spectrometer equipped with a thermoelectric cooled silicon charge-coupled device camera. A set of color filters is used to block the fundamental and other unwanted light. With 810-nm wavelength, band-to-band absorption of fundamental is avoided, and both fundamental and SH can be detected efficiently with silicon detectors, which facilitates alignment and location of the laser spot. Figure \ref{fig:setup}(b) is a microscope photo of a MoS$_2$ flake that is mechanically exfoliated onto a Si/SiO$_2$ (90 nm) substrate.  The region marked with red label 1L is identified as a monolayer according to its optical contrast,\cite{Nanotechnology22125706,small8682,apl96213116}  Raman spectrum,\cite{acsnano42695,b84155413} and photoluminesence spectrum.\cite{l105136805,nl101271} Other regions with few atomic layers, as indicated by the red labels, are assigned according to their relative optical contrasts.

The structure of monolayer MoS$_2$ is schematically shown in Fig.~\ref{fig:setup}(c), where each yellow circle represents two S atoms vertically separated by 0.65 nm, and blue circles indicate plane of Mo atoms located between the two S atomic planes. With the $D_{3h}$ symmetry, the second-order nonlinear susceptibility tensor has nonzero elements of $\chi^{(2)}_{y^{\prime}y^{\prime}y^{\prime}} = - \chi^{(2)}_{y^{\prime}x^{\prime}x^{\prime}} = - \chi^{(2)}_{x^{\prime}x^{\prime}y^{\prime}} = - \chi^{(2)}_{x^{\prime}y^{\prime}x^{\prime}} \equiv \chi^{(2)}$,\cite{bookboyd} where $x^{\prime}y^{\prime}z^{\prime}$ are crystalline coordinates. Here, $x^{\prime}$ is along the armchair direction, which is $30^{\circ}$ from the zig-zag direction, along which the mirror symmetry is broken. In the experiment, the fundamental beam is normal incident (along $-z^{\prime}$) and is linearly polarized along horizontal direction [defined as $x$ in the laboratory coordinates, as shown in Fig.~\ref{fig:setup}(b)]. It is straightforward to show that the parallel ($x$) and perpendicular ($y$) components of SH field are proportional to $\mathrm{sin} 3 \theta$ and $\mathrm{cos} 3 \theta$, respectively, where $\theta$ is the angle between $x$ and $x^{\prime}$. Hence, the power of the two components varies as $P_x \propto\mathrm{sin}^2 3 \theta$ and $P_y \propto \mathrm{cos}^2 3 \theta$, while the total power is independent of $\theta$.

\begin{figure}
\centering
\includegraphics[width=8cm]{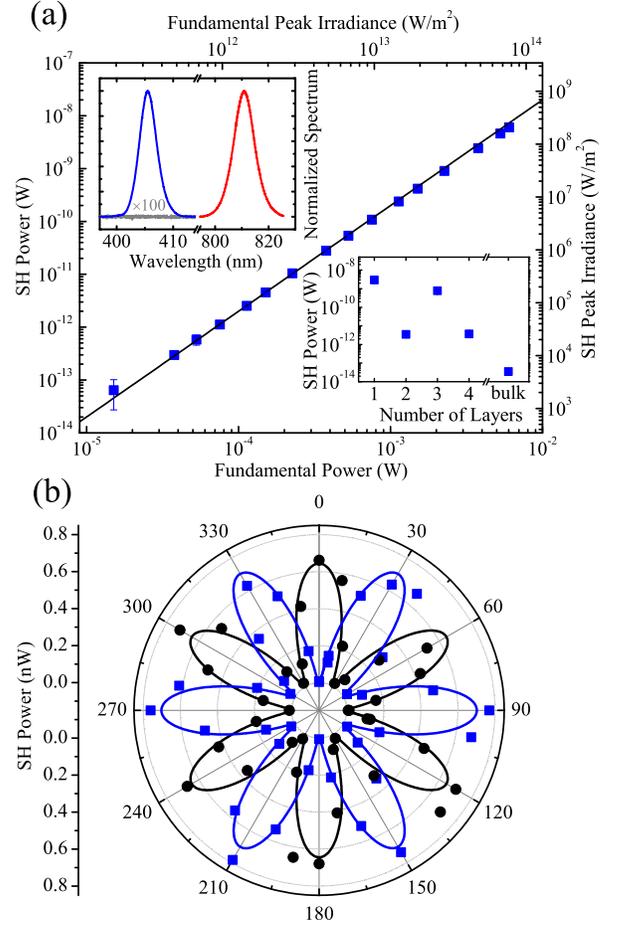}
\caption{Second harmonic generation from mechanically exfoliated MoS$_2$ sample: The upper inset of (a) shows the spectra of second harmonic from the monolayer MoS$_2$ and from bare substrate (gray, multiplied by a factor of 100), as well as the fundamental beams (red). The lower inset shows the second harmonic power measured from regions with different atomic layers. The main panel of (a) shows the power dependence of second harmonic generation, with the solid line indicating the expected quadratic dependence. (b) Power of parallel (blue squares) and perpendicular (black circles) components of second harmonic as a function of $\theta$, the angle between the laboratory and the crystalline coordinates. The blue (black) solid line indicates the expected $\mathrm{sin}^2 3 \theta$ ($\mathrm{cos}^2 3 \theta$) dependence.}
\label{fig:exfoliated}
\end{figure}

In our experiment, we first measure the total SH power with a fundamental power of 4 mW. The upper inset of Fig.~\ref{fig:exfoliated}(a) shows the spectra of the SH (blue) and the fundamental (red), confirming that the former is indeed at half wavelength of the latter. The gray curve is a spectrum (multiplied by a factor of 100) taken under the same conditions but with the laser spot located on bare substrate. Hence, the contrast of monolayer with respect to substrate is at least 10$^4$, which is much higher than linear optical microscopy (about 0.3). The main panel of Fig.~\ref{fig:exfoliated}(a) shows how the SH power varies with the fundamental power. The peak irradiance of fundamental and SH pulses, deduced from the powers, are also plotted for convenience, as top and right axes. The solid line is the expected quadratic dependence for the SHG process. Next, by placing a linear polarizer in front of the spectrometer, we measure $P_x$ and $P_y$ as a function of $\theta$, the angle between $x$ and $x^{\prime}$, by rotating the sample about $z$ axis. Figure~\ref{fig:exfoliated}(b) shows the results, along with the expected $\theta$ dependence (solid lines) from the $D_{3H}$ symmetry.

In order to estimate the magnitude of $\chi^{(2)}$ from the measurement, we model the monolayer as a bulk medium. Since the flake thickness ($d=0.65$ nm) is much smaller than the coherence length, the SHG is not influenced by phase-matching conditions. By solving the coupled-wave equations,\cite{bookboyd} the SH field amplitude of the parallel component
\begin{equation}
\mathcal{E}_{x}=\frac{1}{4}\frac{i2\omega}{2n_{2\omega}c}\chi^{(2)}d \mathcal{E}_\omega^2 \mathrm{sin} 3 \theta,
\label{eq:E3omega}
\end{equation}
where $c$ is the speed of light in a vacuum and $n_{2\omega}$ is the index of refraction at SH, and $\mathcal{E}_{\omega}$ is the fundamental field amplitude. The $\mathcal{E}_{x}$ is related to the irradiance by $I_x = n_{2\omega} \epsilon_0 c \mathcal{E}_x \mathcal{E}_x^* / 2$, which can be calculated from the measured quantity, average power, by considering that $I_x$ is Gaussian in both time and space, with widths (full width at half maxima) of $\tau$ and $W$, respectively. By using $W$ = 2~$\mu$m, $\tau$ = 200 fs, $f$ = 81 MHz, $n_{2\omega} \approx 6.0$,\cite{apl96213116} and reflection coefficient of 0.09 from this multilayer structure, we find that the magnitude of $\chi^{(2)}$ is about $10^{-7}$~m/V. We note that due to the nonlinear nature of this process, such a deduction replies on accurate knowledge on many experimental parameters, such as the shape and duration of the fundamental pulse, the shape and size of the focused fundamental spot at sample, and the relation between the measured spectral counts and the actually SH power. Hence, this value should be viewed as an order-of-magnitude estimate. However, the relative comparison of $\chi^{(2)}$ throughout this paper are not influenced by such uncertainties, and are thus accurate.

Since monolayer MoS$_2$ possesses such a large $\chi^{(2)}$, which vanishes in bulk, it is interesting to study how $\chi^{(2)}$ varies with the number of atomic layers. We measure the total power of SH from different regions of the flake shown in Fig.~\ref{fig:setup}(b), with a fixed fundamental power of 4 mW.  The results are summaries in the lower inset of Fig.~\ref{fig:exfoliated}(a). Since the total power is independent of $\theta$, the measurement is not influenced by potentially different crystal orientations of these regions. We find that $\chi^{(2)}$ of trilayer is about a factor of seven smaller than monolayer, while those of bilayer and quadralayer are about two orders of magnitude smaller than the monolayer. Since flakes with even number of atomic layers possess inversion symmetry, their second-order response should vanish. The smaller but nonzero $\chi^{(2)}$ can be attributed to surface and interface effects. According to this measurement, the contrast of monolayer with respect to bilayer and quaralayer is about 10$^4$. Similar layer-number dependence has also been observed recently in WS$_2$ and WSe$_2$.\cite{arXiv12085864} We also measure a thick flake that can be considered as a bulk. The SH power is about five orders of magnitude smaller than the monolayer, indicating a very small $\chi^{(2)}$, as a result of the inversion symmetry.

The observed SHG can be used to fast and {\it in-situ} characterize atomically thin films of MoS$_2$ and similar materials. Although high quality monolayer MoS$_2$ can be produced by simple mechanical exfoliation\cite{pnas10210451} and identified by optical contrast with certain substrates\cite{Nanotechnology22125706,small8682} and Raman spectroscopy,\cite{acsnano42695,b84155413} applications of this material rely on development of scalable techniques. Following initial works of mechanical exfoliation, other top-down methods with better potential for large-scale production have been developed, such as lithium ion exfoliation\cite{nl115111,acie5011093,mrs10287,jacs1255998,acie519052} and ultrasonic exfoliation in liquids.\cite{s331568,am233944,jpcc11611393,acie5010839,acsnano63468,cm233879,cm242414} Promising progresses have also been made in developing bottom-up methods, including hydrothermal synthesis\cite{jssc159170,cl30772} and CVD on insulating substrates\cite{am242320,small8966,nl121538,nanoscale4461} and graphene.\cite{nl122784} However, one significant obstacle is the lack of techniques for fast and {\it in-situ} sample characterization. For example, thin films of MoS$_2$ fabricated by these techniques are polycrystalline. They are composed of single-crystalline domains with random crystal orientations and are separated by grain boundaries, which severely limit performance of the films, especially their conductivity and mechanical strength. However, it is difficult to locate the grain boundaries and monitor size of these domains {\it in-situ}. 

\begin{figure}
\centering
\includegraphics[width=8cm]{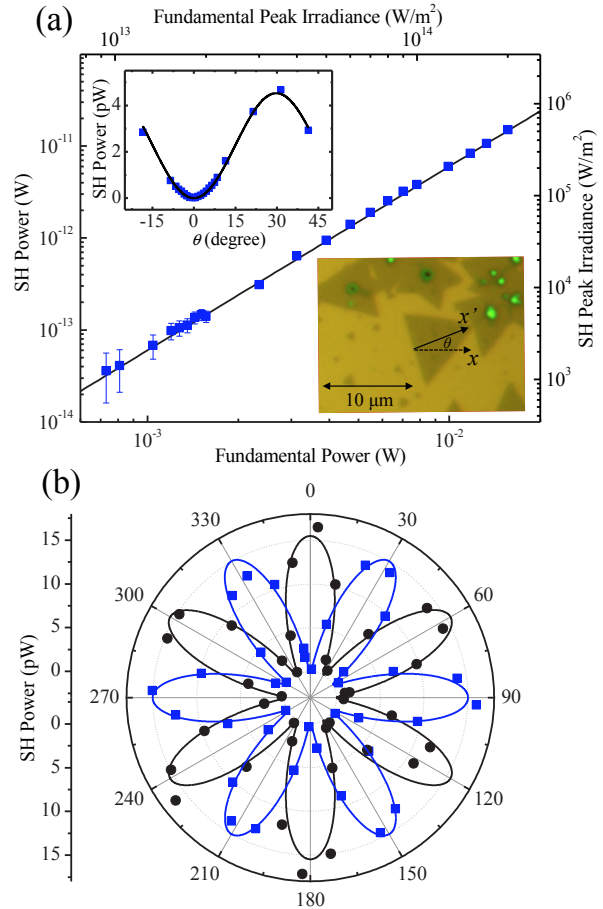}
\caption{Second harmonic generation from a triangular monolayer MoS$_2$ flake grown by CVD, as shown in the lower inset of (a). The main panel of (a) shows the power dependence of second harmonic generation. The solid line indicates the expected quadratic dependence. Panel (b) shows angular dependence of the parallel (blue squares) and perpendicular (black circles) components of second harmonic, along with the expected dependence (solid lines). The upper inset of (a) shows a separate measurement of the parallel component with a finer step size near $\theta = 0^{\circ}$.}
\label{fig:triangle}
\end{figure}

The lower inset of Fig.~\ref{fig:triangle}(a) shows a microscope photo of some triangular monolayer MoS$_2$ flakes on a Si/SiO$_2$(280 nm) substrate fabricated by CVD. The samples were prepared using MoO$_3$ and sublimated sulfur as precursors. MoO$_3$ is positioned close to the designated growth substrate at the center of the furnace, while sublimated sulfur is positioned upstream at a zone where evaporation starts at 750$^\circ$C. The reaction of the precursors at 850$^\circ$C in a furnace flushed with nitrogen results in nucleation of single crystalline domains. The density of nucleation and samples sizes can be controlled by monitoring the pressure and the closely related sulfur concentration in the chamber. By maintaining a positive pressure in the range of 5 - 20 KPa, MoS$_2$ domains with sizes in the range of 10 - 40 $\mu$m are synthesized, with a ramping time of 60 to 90 minutes and 10 minutes at the reaction temperature.\cite{arxiv13012812} 

The main panel of Fig. \ref{fig:triangle}(a) shows the quadratic power dependence of SHG, similar to Fig.~\ref{fig:exfoliated}(a), measured from the well-separated flake on which the crystalline and laboratory coordinates ($x^{\prime}$ and $x$, respectively) are plotted [the lower inset of Fig.~\ref{fig:triangle}(a)]. By rotating the sample, we measure $P_x$ and $P_y$ as a function of $\theta$, as shown in Fig.~\ref{fig:triangle}(b). The results are similar to the exfoliated sample shown in Fig.~\ref{fig:exfoliated}(b). A separate measurement of the parallel component with finer resolution near $\theta = 0^{\circ}$, shown in the upper inset of Fig.~\ref{fig:triangle}(a), confirms that the minimal parallel component occurs precisely at $\theta = 0^{\circ}$. The edges of these triangular flakes are expected to be along zig-zag directions since these are lowest energy configurations.\cite{nn253,cl6495} The maximum parallel component of SH should occurs when the fundamental is polarized along the zig-zag direction, which is consistent our observation. Hence, the SHG further confirms that the direction of the edges is zig-zag. We repeat the measurement with several other similar triangular flakes, and obtained the same result. Such an established relation also allows us to determine the crystal orientation of the mechanically exfoliated sample shown in Fig.~\ref{fig:setup}(b): that is, the armchair direction of the 1L region is horizontal, and its lattice orientation is as shown in Fig.~\ref{fig:setup}(c). From the strength of the SH, we deduce a $\chi^{(2)} \approx 5 \times 10^{-9}$ m/V, which is about a factor of 20 smaller than the mechanically exfoliated flake. 

\begin{figure}
\centering
\includegraphics[width=8.5cm]{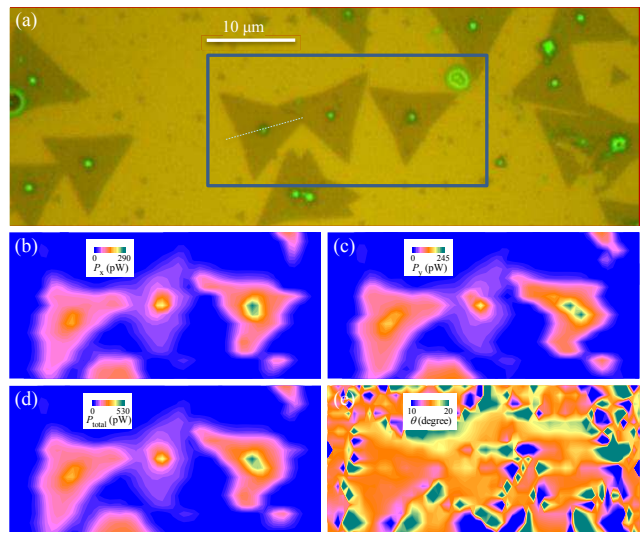}
\caption{(a) Optical microscopy photo of a region of substrate containing flakes grown by CVD. (b) and (c): Maps of $P_x$ and $P_y$ over the region indicated by the box in (a). (d) Map of the total power, $P_x + P_y$. (e) Map of $\theta$ calculated from (b) and (c).}
\label{fig:2D}
\end{figure}

Figure \ref{fig:2D} summarizes our proof-of-principle demonstration of a polarization-revolved SH microscopy. We study a region on the substrate with quasi-continuos films, as shown in Fig.~\ref{fig:2D}(a). It contains a high density but still separated and randomly oriented triangular flakes, so that we can correlate domains observed in SHG to the actual regions. In this measurement, we scan a 20-mW fundamental spot across the region indicated by the box in Fig.~\ref{fig:2D}(a), and detect the powers of the parallel and perpendicular components of the SH, as shown in Figs.~\ref{fig:2D}(b) and (c), respectively. The errors in these scans are below 2 pW, or smaller than 1\% of the maximum signal. Figure \ref{fig:2D}(d) shows the total power, obtained by adding (b) and (c). From (b) and (c), we calculate the angle by using $\theta = (1/3) \mathrm{tan}^{-1} \sqrt{P_x / P_y}$, as shown in (e). The uncertainty on the angle is below 1$^{\circ}$.

The combination of linear and nonlinear optical microscopy can provide valuable information on polycrystalline thin films grown by CVD. First, the greenish dots in Fig.~\ref{fig:2D}(a) indicate that at the central area of some flakes, a second (or even third) layer is grown. Second harmonic images show that these areas have {\it higher} SH power. In bilayer MoS$_2$ exfoliated from 2$H$-stacked crystals, the two layers are inversely oriented so that the bilayer possesses inversion symmetry. Hence, it second-order response should vanish, as confirmed in Fig.~\ref{fig:exfoliated}(a). The higher SH power observed from multilayer regions of CVD-grown flakes indicates that these multilayers are not 2$H$-stacked. This is similar to multilayer graphene grown by CVD. Clearly, the SH microscopy is capable of probing relative orientations among multilayers of MoS$_2$. Second, panel (e) shows that $\theta$ is uniform over the left flake, which is about $15^{\circ}$. This is consistent with the shape observed in (a) (white dashed line). With further growth time, this flake will merge with other flakes to form a continuous polycrystalline film. Linear optical microscopy would not allow identification of each single-crystalline domains. However the $\theta$ map can still distinguish these domains, locate their boundaries, and measure their sizes. Third, the shapes of these flakes are irregular in the SH maps. Especially, the parallel and perpendicular components have different edge shapes. This can be attributed to the roughness on the edges and different termination configurations on the edges. Although further characterizations are needed to correlate the microscopic structure on the edges to the SH power, this observation illustrates the potential of using SHG to study these edge structures. Finally, although the three flakes look similar in (a), the SH power are different, and the $\theta$ of the middle and right flakes are irregular. This illustrates that the SH microscopy can show different properties and qualities of the flakes that the linear optical microscopy cannot. However, further studies are needed to correlate SHG to these specific sample characteristics.

In summary, we have observed strong second harmonic generation in monolayer MoS$_2$ fabricated by mechanical exfoliation and CVD, and performed a proof-of-principle second harmonic microscopy measurement. Our results show that such a nonlinear optical effect can be used to fast and non-invasively characterize atomically thin films of MoS$_2$ and other similar materials. Compared to linear optical microscopy, the contrast is enhanced by at least four orders of magnitude. Unlike linear optical microscopy that replies on light interference from carefully designed multilayer substrates, such a nonlinear optical microscopy can be applied to any substrates with weak second-order nonlinearity, such as silicon and glass. Although Raman microscopy has also been used to identify monolayer MoS$_2$, the Raman shift often depends on substrates, and the contrast is relatively low. In addition to these advantages in identifying monolayers, the second harmonic microscopy can probe crystal orientation, single-crystal domain size, and layer stacking.

HZ acknowledges support from the US National Science Foundation under Awards No. DMR-0954486 and No. EPS-0903806, and matching support from the State of Kansas through Kansas Technology Enterprise Corporation. JL acknowledges support from Welch Foundation (C-1716).


\begin{thebibliography}{53}%
\makeatletter
\providecommand \@ifxundefined [1]{%
 \@ifx{#1\undefined}
}%
\providecommand \@ifnum [1]{%
 \ifnum #1\expandafter \@firstoftwo
 \else \expandafter \@secondoftwo
 \fi
}%
\providecommand \@ifx [1]{%
 \ifx #1\expandafter \@firstoftwo
 \else \expandafter \@secondoftwo
 \fi
}%
\providecommand \natexlab [1]{#1}%
\providecommand \enquote  [1]{``#1''}%
\providecommand \bibnamefont  [1]{#1}%
\providecommand \bibfnamefont [1]{#1}%
\providecommand \citenamefont [1]{#1}%
\providecommand \href@noop [0]{\@secondoftwo}%
\providecommand \href [0]{\begingroup \@sanitize@url \@href}%
\providecommand \@href[1]{\@@startlink{#1}\@@href}%
\providecommand \@@href[1]{\endgroup#1\@@endlink}%
\providecommand \@sanitize@url [0]{\catcode `\\12\catcode `\$12\catcode
  `\&12\catcode `\#12\catcode `\^12\catcode `\_12\catcode `\%12\relax}%
\providecommand \@@startlink[1]{}%
\providecommand \@@endlink[0]{}%
\providecommand \url  [0]{\begingroup\@sanitize@url \@url }%
\providecommand \@url [1]{\endgroup\@href {#1}{\urlprefix }}%
\providecommand \urlprefix  [0]{URL }%
\providecommand \Eprint [0]{\href }%
\providecommand \doibase [0]{http://dx.doi.org/}%
\providecommand \selectlanguage [0]{\@gobble}%
\providecommand \bibinfo  [0]{\@secondoftwo}%
\providecommand \bibfield  [0]{\@secondoftwo}%
\providecommand \translation [1]{[#1]}%
\providecommand \BibitemOpen [0]{}%
\providecommand \bibitemStop [0]{}%
\providecommand \bibitemNoStop [0]{.\EOS\space}%
\providecommand \EOS [0]{\spacefactor3000\relax}%
\providecommand \BibitemShut  [1]{\csname bibitem#1\endcsname}%
\let\auto@bib@innerbib\@empty
\bibitem [{\citenamefont {Wang}\ \emph
  {et~al.}(2012{\natexlab{a}})\citenamefont {Wang}, \citenamefont
  {Kalantar-Zadeh}, \citenamefont {Kis}, \citenamefont {Coleman},\ and\
  \citenamefont {Strano}}]{nn7699}%
  \BibitemOpen
  \bibfield  {author} {\bibinfo {author} {\bibfnamefont {Q.~H.}\ \bibnamefont
  {Wang}}, \bibinfo {author} {\bibfnamefont {K.}~\bibnamefont
  {Kalantar-Zadeh}}, \bibinfo {author} {\bibfnamefont {A.}~\bibnamefont {Kis}},
  \bibinfo {author} {\bibfnamefont {J.~N.}\ \bibnamefont {Coleman}}, \ and\
  \bibinfo {author} {\bibfnamefont {M.~S.}\ \bibnamefont {Strano}},\
  }\href@noop {} {\bibfield  {journal} {\bibinfo  {journal} {Nat.
  Nanotechnol.}\ }\textbf {\bibinfo {volume} {7}},\ \bibinfo {pages} {699}
  (\bibinfo {year} {2012}{\natexlab{a}})}\BibitemShut {NoStop}%
\bibitem [{\citenamefont {Mak}\ \emph {et~al.}(2010)\citenamefont {Mak},
  \citenamefont {Lee}, \citenamefont {Hone}, \citenamefont {Shan},\ and\
  \citenamefont {Heinz}}]{l105136805}%
  \BibitemOpen
  \bibfield  {author} {\bibinfo {author} {\bibfnamefont {K.~F.}\ \bibnamefont
  {Mak}}, \bibinfo {author} {\bibfnamefont {C.}~\bibnamefont {Lee}}, \bibinfo
  {author} {\bibfnamefont {J.}~\bibnamefont {Hone}}, \bibinfo {author}
  {\bibfnamefont {J.}~\bibnamefont {Shan}}, \ and\ \bibinfo {author}
  {\bibfnamefont {T.~F.}\ \bibnamefont {Heinz}},\ }\href@noop {} {\bibfield
  {journal} {\bibinfo  {journal} {Phys. Rev. Lett.}\ }\textbf {\bibinfo
  {volume} {105}},\ \bibinfo {pages} {136805} (\bibinfo {year}
  {2010})}\BibitemShut {NoStop}%
\bibitem [{\citenamefont {Splendiani}\ \emph {et~al.}(2010)\citenamefont
  {Splendiani}, \citenamefont {Sun}, \citenamefont {Zhang}, \citenamefont {Li},
  \citenamefont {Kim}, \citenamefont {Chim}, \citenamefont {Galli},\ and\
  \citenamefont {Wang}}]{nl101271}%
  \BibitemOpen
  \bibfield  {author} {\bibinfo {author} {\bibfnamefont {A.}~\bibnamefont
  {Splendiani}}, \bibinfo {author} {\bibfnamefont {L.}~\bibnamefont {Sun}},
  \bibinfo {author} {\bibfnamefont {Y.}~\bibnamefont {Zhang}}, \bibinfo
  {author} {\bibfnamefont {T.}~\bibnamefont {Li}}, \bibinfo {author}
  {\bibfnamefont {J.}~\bibnamefont {Kim}}, \bibinfo {author} {\bibfnamefont
  {C.~Y.}\ \bibnamefont {Chim}}, \bibinfo {author} {\bibfnamefont
  {G.}~\bibnamefont {Galli}}, \ and\ \bibinfo {author} {\bibfnamefont
  {F.}~\bibnamefont {Wang}},\ }\href@noop {} {\bibfield  {journal} {\bibinfo
  {journal} {Nano Lett.}\ }\textbf {\bibinfo {volume} {10}},\ \bibinfo {pages}
  {1271} (\bibinfo {year} {2010})}\BibitemShut {NoStop}%
\bibitem [{\citenamefont {Kuc}\ \emph {et~al.}(2011)\citenamefont {Kuc},
  \citenamefont {Zibouche},\ and\ \citenamefont {Heine}}]{b83245213}%
  \BibitemOpen
  \bibfield  {author} {\bibinfo {author} {\bibfnamefont {A.}~\bibnamefont
  {Kuc}}, \bibinfo {author} {\bibfnamefont {N.}~\bibnamefont {Zibouche}}, \
  and\ \bibinfo {author} {\bibfnamefont {T.}~\bibnamefont {Heine}},\
  }\href@noop {} {\bibfield  {journal} {\bibinfo  {journal} {Phys. Rev. B}\
  }\textbf {\bibinfo {volume} {83}},\ \bibinfo {pages} {245213} (\bibinfo
  {year} {2011})}\BibitemShut {NoStop}%
\bibitem [{\citenamefont {Liu}\ \emph {et~al.}(2011)\citenamefont {Liu},
  \citenamefont {Kumar}, \citenamefont {Ouyang},\ and\ \citenamefont
  {Guo}}]{ieeeted583042}%
  \BibitemOpen
  \bibfield  {author} {\bibinfo {author} {\bibfnamefont {L.}~\bibnamefont
  {Liu}}, \bibinfo {author} {\bibfnamefont {S.~B.}\ \bibnamefont {Kumar}},
  \bibinfo {author} {\bibfnamefont {Y.}~\bibnamefont {Ouyang}}, \ and\ \bibinfo
  {author} {\bibfnamefont {J.}~\bibnamefont {Guo}},\ }\href@noop {} {\bibfield
  {journal} {\bibinfo  {journal} {IEEE Trans. on Electron Devices}\ }\textbf
  {\bibinfo {volume} {58}},\ \bibinfo {pages} {3042} (\bibinfo {year}
  {2011})}\BibitemShut {NoStop}%
\bibitem [{\citenamefont {Yoon}\ \emph {et~al.}(2011)\citenamefont {Yoon},
  \citenamefont {Ganapathi},\ and\ \citenamefont {Salahuddin}}]{nl113768}%
  \BibitemOpen
  \bibfield  {author} {\bibinfo {author} {\bibfnamefont {Y.}~\bibnamefont
  {Yoon}}, \bibinfo {author} {\bibfnamefont {K.}~\bibnamefont {Ganapathi}}, \
  and\ \bibinfo {author} {\bibfnamefont {S.}~\bibnamefont {Salahuddin}},\
  }\href@noop {} {\bibfield  {journal} {\bibinfo  {journal} {Nano Lett.}\
  }\textbf {\bibinfo {volume} {11}},\ \bibinfo {pages} {3768} (\bibinfo {year}
  {2011})}\BibitemShut {NoStop}%
\bibitem [{\citenamefont {Radisavljevic}\ \emph
  {et~al.}(2011{\natexlab{a}})\citenamefont {Radisavljevic}, \citenamefont
  {Radenovic}, \citenamefont {Brivio}, \citenamefont {Giacometti},\ and\
  \citenamefont {Kis}}]{nn6147}%
  \BibitemOpen
  \bibfield  {author} {\bibinfo {author} {\bibfnamefont {B.}~\bibnamefont
  {Radisavljevic}}, \bibinfo {author} {\bibfnamefont {A.}~\bibnamefont
  {Radenovic}}, \bibinfo {author} {\bibfnamefont {J.}~\bibnamefont {Brivio}},
  \bibinfo {author} {\bibfnamefont {V.}~\bibnamefont {Giacometti}}, \ and\
  \bibinfo {author} {\bibfnamefont {A.}~\bibnamefont {Kis}},\ }\href@noop {}
  {\bibfield  {journal} {\bibinfo  {journal} {Nat. Nanotechnol.}\ }\textbf
  {\bibinfo {volume} {6}},\ \bibinfo {pages} {147} (\bibinfo {year}
  {2011}{\natexlab{a}})}\BibitemShut {NoStop}%
\bibitem [{\citenamefont {Radisavljevic}\ \emph
  {et~al.}(2011{\natexlab{b}})\citenamefont {Radisavljevic}, \citenamefont
  {Whitwick},\ and\ \citenamefont {Kis}}]{acsnano59934}%
  \BibitemOpen
  \bibfield  {author} {\bibinfo {author} {\bibfnamefont {B.}~\bibnamefont
  {Radisavljevic}}, \bibinfo {author} {\bibfnamefont {M.~B.}\ \bibnamefont
  {Whitwick}}, \ and\ \bibinfo {author} {\bibfnamefont {A.}~\bibnamefont
  {Kis}},\ }\href@noop {} {\bibfield  {journal} {\bibinfo  {journal} {ACS
  Nano}\ }\textbf {\bibinfo {volume} {5}},\ \bibinfo {pages} {9934} (\bibinfo
  {year} {2011}{\natexlab{b}})}\BibitemShut {NoStop}%
\bibitem [{\citenamefont {Wang}\ \emph
  {et~al.}(2012{\natexlab{b}})\citenamefont {Wang}, \citenamefont {Yu},
  \citenamefont {Lee}, \citenamefont {Shi}, \citenamefont {Hsu}, \citenamefont
  {Chin}, \citenamefont {Li}, \citenamefont {Dubey}, \citenamefont {Kong},\
  and\ \citenamefont {Palacios}}]{nl124674}%
  \BibitemOpen
  \bibfield  {author} {\bibinfo {author} {\bibfnamefont {H.}~\bibnamefont
  {Wang}}, \bibinfo {author} {\bibfnamefont {L.~L.}\ \bibnamefont {Yu}},
  \bibinfo {author} {\bibfnamefont {Y.~H.}\ \bibnamefont {Lee}}, \bibinfo
  {author} {\bibfnamefont {Y.~M.}\ \bibnamefont {Shi}}, \bibinfo {author}
  {\bibfnamefont {A.}~\bibnamefont {Hsu}}, \bibinfo {author} {\bibfnamefont
  {M.~L.}\ \bibnamefont {Chin}}, \bibinfo {author} {\bibfnamefont {L.~J.}\
  \bibnamefont {Li}}, \bibinfo {author} {\bibfnamefont {M.}~\bibnamefont
  {Dubey}}, \bibinfo {author} {\bibfnamefont {J.}~\bibnamefont {Kong}}, \ and\
  \bibinfo {author} {\bibfnamefont {T.}~\bibnamefont {Palacios}},\ }\href@noop
  {} {\bibfield  {journal} {\bibinfo  {journal} {Nano Lett.}\ }\textbf
  {\bibinfo {volume} {12}},\ \bibinfo {pages} {4674} (\bibinfo {year}
  {2012}{\natexlab{b}})}\BibitemShut {NoStop}%
\bibitem [{\citenamefont {Zhang}\ \emph {et~al.}(2012)\citenamefont {Zhang},
  \citenamefont {Ye}, \citenamefont {Matsuhashi},\ and\ \citenamefont
  {Iwasa}}]{nl121136}%
  \BibitemOpen
  \bibfield  {author} {\bibinfo {author} {\bibfnamefont {Y.}~\bibnamefont
  {Zhang}}, \bibinfo {author} {\bibfnamefont {J.}~\bibnamefont {Ye}}, \bibinfo
  {author} {\bibfnamefont {Y.}~\bibnamefont {Matsuhashi}}, \ and\ \bibinfo
  {author} {\bibfnamefont {Y.}~\bibnamefont {Iwasa}},\ }\href@noop {}
  {\bibfield  {journal} {\bibinfo  {journal} {Nano Lett.}\ }\textbf {\bibinfo
  {volume} {12}},\ \bibinfo {pages} {1136} (\bibinfo {year}
  {2012})}\BibitemShut {NoStop}%
\bibitem [{\citenamefont {Bertolazzi}\ \emph {et~al.}(2011)\citenamefont
  {Bertolazzi}, \citenamefont {Brivio},\ and\ \citenamefont
  {Kis}}]{acsnano59703}%
  \BibitemOpen
  \bibfield  {author} {\bibinfo {author} {\bibfnamefont {S.}~\bibnamefont
  {Bertolazzi}}, \bibinfo {author} {\bibfnamefont {J.}~\bibnamefont {Brivio}},
  \ and\ \bibinfo {author} {\bibfnamefont {A.}~\bibnamefont {Kis}},\
  }\href@noop {} {\bibfield  {journal} {\bibinfo  {journal} {ACS Nano}\
  }\textbf {\bibinfo {volume} {5}},\ \bibinfo {pages} {9703} (\bibinfo {year}
  {2011})}\BibitemShut {NoStop}%
\bibitem [{\citenamefont {He}\ \emph {et~al.}(2012)\citenamefont {He},
  \citenamefont {Zeng}, \citenamefont {Yin}, \citenamefont {Li}, \citenamefont
  {Wu}, \citenamefont {Huang},\ and\ \citenamefont {Zhang}}]{small82994}%
  \BibitemOpen
  \bibfield  {author} {\bibinfo {author} {\bibfnamefont {Q.~Y.}\ \bibnamefont
  {He}}, \bibinfo {author} {\bibfnamefont {Z.~Y.}\ \bibnamefont {Zeng}},
  \bibinfo {author} {\bibfnamefont {Z.~Y.}\ \bibnamefont {Yin}}, \bibinfo
  {author} {\bibfnamefont {H.}~\bibnamefont {Li}}, \bibinfo {author}
  {\bibfnamefont {S.~X.}\ \bibnamefont {Wu}}, \bibinfo {author} {\bibfnamefont
  {X.}~\bibnamefont {Huang}}, \ and\ \bibinfo {author} {\bibfnamefont
  {H.}~\bibnamefont {Zhang}},\ }\href@noop {} {\bibfield  {journal} {\bibinfo
  {journal} {Small}\ }\textbf {\bibinfo {volume} {8}},\ \bibinfo {pages} {2994}
  (\bibinfo {year} {2012})}\BibitemShut {NoStop}%
\bibitem [{\citenamefont {Pu}\ \emph {et~al.}(2012)\citenamefont {Pu},
  \citenamefont {Yomogida}, \citenamefont {Liu}, \citenamefont {Li},
  \citenamefont {Iwasa},\ and\ \citenamefont {Takenobu}}]{nl124013}%
  \BibitemOpen
  \bibfield  {author} {\bibinfo {author} {\bibfnamefont {J.}~\bibnamefont
  {Pu}}, \bibinfo {author} {\bibfnamefont {Y.}~\bibnamefont {Yomogida}},
  \bibinfo {author} {\bibfnamefont {K.~K.}\ \bibnamefont {Liu}}, \bibinfo
  {author} {\bibfnamefont {L.~J.}\ \bibnamefont {Li}}, \bibinfo {author}
  {\bibfnamefont {Y.}~\bibnamefont {Iwasa}}, \ and\ \bibinfo {author}
  {\bibfnamefont {T.}~\bibnamefont {Takenobu}},\ }\href@noop {} {\bibfield
  {journal} {\bibinfo  {journal} {Nano Lett.}\ }\textbf {\bibinfo {volume}
  {12}},\ \bibinfo {pages} {4013} (\bibinfo {year} {2012})}\BibitemShut
  {NoStop}%
\bibitem [{\citenamefont {Mak}\ \emph {et~al.}(2013)\citenamefont {Mak},
  \citenamefont {He}, \citenamefont {Lee}, \citenamefont {Lee}, \citenamefont
  {Hone}, \citenamefont {Heinz},\ and\ \citenamefont {Shan}}]{nm12207}%
  \BibitemOpen
  \bibfield  {author} {\bibinfo {author} {\bibfnamefont {K.~F.}\ \bibnamefont
  {Mak}}, \bibinfo {author} {\bibfnamefont {K.}~\bibnamefont {He}}, \bibinfo
  {author} {\bibfnamefont {C.}~\bibnamefont {Lee}}, \bibinfo {author}
  {\bibfnamefont {G.~H.}\ \bibnamefont {Lee}}, \bibinfo {author} {\bibfnamefont
  {J.}~\bibnamefont {Hone}}, \bibinfo {author} {\bibfnamefont {T.~F.}\
  \bibnamefont {Heinz}}, \ and\ \bibinfo {author} {\bibfnamefont
  {J.}~\bibnamefont {Shan}},\ }\href@noop {} {\bibfield  {journal} {\bibinfo
  {journal} {Nat. Mater.}\ }\textbf {\bibinfo {volume} {12}},\ \bibinfo {pages}
  {207} (\bibinfo {year} {2013})}\BibitemShut {NoStop}%
\bibitem [{\citenamefont {Yin}\ \emph {et~al.}(2012)\citenamefont {Yin},
  \citenamefont {Li}, \citenamefont {Li}, \citenamefont {Jiang}, \citenamefont
  {Shi}, \citenamefont {Sun}, \citenamefont {Lu}, \citenamefont {Zhang},
  \citenamefont {Chen},\ and\ \citenamefont {Zhang}}]{acsnano674}%
  \BibitemOpen
  \bibfield  {author} {\bibinfo {author} {\bibfnamefont {Z.}~\bibnamefont
  {Yin}}, \bibinfo {author} {\bibfnamefont {H.}~\bibnamefont {Li}}, \bibinfo
  {author} {\bibfnamefont {H.}~\bibnamefont {Li}}, \bibinfo {author}
  {\bibfnamefont {L.}~\bibnamefont {Jiang}}, \bibinfo {author} {\bibfnamefont
  {Y.}~\bibnamefont {Shi}}, \bibinfo {author} {\bibfnamefont {Y.}~\bibnamefont
  {Sun}}, \bibinfo {author} {\bibfnamefont {G.}~\bibnamefont {Lu}}, \bibinfo
  {author} {\bibfnamefont {Q.}~\bibnamefont {Zhang}}, \bibinfo {author}
  {\bibfnamefont {X.}~\bibnamefont {Chen}}, \ and\ \bibinfo {author}
  {\bibfnamefont {H.}~\bibnamefont {Zhang}},\ }\href@noop {} {\bibfield
  {journal} {\bibinfo  {journal} {ACS Nano}\ }\textbf {\bibinfo {volume} {6}},\
  \bibinfo {pages} {74} (\bibinfo {year} {2012})}\BibitemShut {NoStop}%
\bibitem [{\citenamefont {Lee}\ \emph {et~al.}(2012{\natexlab{a}})\citenamefont
  {Lee}, \citenamefont {Min}, \citenamefont {Chang}, \citenamefont {Park},
  \citenamefont {Nam}, \citenamefont {Kim}, \citenamefont {Kim}, \citenamefont
  {Ryu},\ and\ \citenamefont {Im}}]{nl123695}%
  \BibitemOpen
  \bibfield  {author} {\bibinfo {author} {\bibfnamefont {H.~S.}\ \bibnamefont
  {Lee}}, \bibinfo {author} {\bibfnamefont {S.~W.}\ \bibnamefont {Min}},
  \bibinfo {author} {\bibfnamefont {Y.~G.}\ \bibnamefont {Chang}}, \bibinfo
  {author} {\bibfnamefont {M.~K.}\ \bibnamefont {Park}}, \bibinfo {author}
  {\bibfnamefont {T.}~\bibnamefont {Nam}}, \bibinfo {author} {\bibfnamefont
  {H.}~\bibnamefont {Kim}}, \bibinfo {author} {\bibfnamefont {J.~H.}\
  \bibnamefont {Kim}}, \bibinfo {author} {\bibfnamefont {S.}~\bibnamefont
  {Ryu}}, \ and\ \bibinfo {author} {\bibfnamefont {S.}~\bibnamefont {Im}},\
  }\href@noop {} {\bibfield  {journal} {\bibinfo  {journal} {Nano Lett.}\
  }\textbf {\bibinfo {volume} {12}},\ \bibinfo {pages} {3695} (\bibinfo {year}
  {2012}{\natexlab{a}})}\BibitemShut {NoStop}%
\bibitem [{\citenamefont {Shanmugam}\ \emph {et~al.}(2012)\citenamefont
  {Shanmugam}, \citenamefont {Bansal}, \citenamefont {Durcan},\ and\
  \citenamefont {Yu}}]{apl100153901}%
  \BibitemOpen
  \bibfield  {author} {\bibinfo {author} {\bibfnamefont {M.}~\bibnamefont
  {Shanmugam}}, \bibinfo {author} {\bibfnamefont {T.}~\bibnamefont {Bansal}},
  \bibinfo {author} {\bibfnamefont {C.~A.}\ \bibnamefont {Durcan}}, \ and\
  \bibinfo {author} {\bibfnamefont {B.}~\bibnamefont {Yu}},\ }\href@noop {}
  {\bibfield  {journal} {\bibinfo  {journal} {Appl. Phys. Lett.}\ }\textbf
  {\bibinfo {volume} {100}},\ \bibinfo {pages} {153901} (\bibinfo {year}
  {2012})}\BibitemShut {NoStop}%
\bibitem [{\citenamefont {Boyd}(2008)}]{bookboyd}%
  \BibitemOpen
  \bibfield  {author} {\bibinfo {author} {\bibfnamefont {R.~W.}\ \bibnamefont
  {Boyd}},\ }\href@noop {} {\emph {\bibinfo {title} {Nonlinear Optics}}},\
  \bibinfo {edition} {3rd}\ ed.\ (\bibinfo  {publisher} {Academy Press},\
  \bibinfo {address} {San Diego, USA},\ \bibinfo {year} {2008})\BibitemShut
  {NoStop}%
\bibitem [{\citenamefont {Wagoner}\ \emph {et~al.}(1998)\citenamefont
  {Wagoner}, \citenamefont {Persans}, \citenamefont {{Van Wagenen}},\ and\
  \citenamefont {Korenowski}}]{josab151017}%
  \BibitemOpen
  \bibfield  {author} {\bibinfo {author} {\bibfnamefont {G.~A.}\ \bibnamefont
  {Wagoner}}, \bibinfo {author} {\bibfnamefont {P.~D.}\ \bibnamefont
  {Persans}}, \bibinfo {author} {\bibfnamefont {E.~A.}\ \bibnamefont {{Van
  Wagenen}}}, \ and\ \bibinfo {author} {\bibfnamefont {G.~M.}\ \bibnamefont
  {Korenowski}},\ }\href@noop {} {\bibfield  {journal} {\bibinfo  {journal} {J.
  Opt. Soc. Am. B}\ }\textbf {\bibinfo {volume} {15}},\ \bibinfo {pages} {1017}
  (\bibinfo {year} {1998})}\BibitemShut {NoStop}%
\bibitem [{\citenamefont {Cao}\ \emph {et~al.}(2012)\citenamefont {Cao},
  \citenamefont {Wang}, \citenamefont {Han}, \citenamefont {Ye}, \citenamefont
  {Zhu}, \citenamefont {Shi}, \citenamefont {Niu}, \citenamefont {Tan},
  \citenamefont {Wang}, \citenamefont {Liu},\ and\ \citenamefont
  {Feng}}]{nc3887}%
  \BibitemOpen
  \bibfield  {author} {\bibinfo {author} {\bibfnamefont {T.}~\bibnamefont
  {Cao}}, \bibinfo {author} {\bibfnamefont {G.}~\bibnamefont {Wang}}, \bibinfo
  {author} {\bibfnamefont {W.}~\bibnamefont {Han}}, \bibinfo {author}
  {\bibfnamefont {H.}~\bibnamefont {Ye}}, \bibinfo {author} {\bibfnamefont
  {C.}~\bibnamefont {Zhu}}, \bibinfo {author} {\bibfnamefont {J.}~\bibnamefont
  {Shi}}, \bibinfo {author} {\bibfnamefont {Q.}~\bibnamefont {Niu}}, \bibinfo
  {author} {\bibfnamefont {P.}~\bibnamefont {Tan}}, \bibinfo {author}
  {\bibfnamefont {E.}~\bibnamefont {Wang}}, \bibinfo {author} {\bibfnamefont
  {B.}~\bibnamefont {Liu}}, \ and\ \bibinfo {author} {\bibfnamefont
  {J.}~\bibnamefont {Feng}},\ }\href@noop {} {\bibfield  {journal} {\bibinfo
  {journal} {Nat. Commun.}\ }\textbf {\bibinfo {volume} {3}},\ \bibinfo {pages}
  {887} (\bibinfo {year} {2012})}\BibitemShut {NoStop}%
\bibitem [{\citenamefont {Zeng}\ \emph {et~al.}(2012)\citenamefont {Zeng},
  \citenamefont {Dai}, \citenamefont {Yao}, \citenamefont {Xiao},\ and\
  \citenamefont {Cui}}]{nnano7490}%
  \BibitemOpen
  \bibfield  {author} {\bibinfo {author} {\bibfnamefont {H.}~\bibnamefont
  {Zeng}}, \bibinfo {author} {\bibfnamefont {J.}~\bibnamefont {Dai}}, \bibinfo
  {author} {\bibfnamefont {W.}~\bibnamefont {Yao}}, \bibinfo {author}
  {\bibfnamefont {D.}~\bibnamefont {Xiao}}, \ and\ \bibinfo {author}
  {\bibfnamefont {X.}~\bibnamefont {Cui}},\ }\href@noop {} {\bibfield
  {journal} {\bibinfo  {journal} {Nat. Nano.}\ }\textbf {\bibinfo {volume}
  {7}},\ \bibinfo {pages} {490} (\bibinfo {year} {2012})}\BibitemShut {NoStop}%
\bibitem [{\citenamefont {Mak}\ \emph {et~al.}(2012)\citenamefont {Mak},
  \citenamefont {He}, \citenamefont {Shan},\ and\ \citenamefont
  {Heinz}}]{nnano7494}%
  \BibitemOpen
  \bibfield  {author} {\bibinfo {author} {\bibfnamefont {K.~F.}\ \bibnamefont
  {Mak}}, \bibinfo {author} {\bibfnamefont {K.}~\bibnamefont {He}}, \bibinfo
  {author} {\bibfnamefont {J.}~\bibnamefont {Shan}}, \ and\ \bibinfo {author}
  {\bibfnamefont {T.~F.}\ \bibnamefont {Heinz}},\ }\href@noop {} {\bibfield
  {journal} {\bibinfo  {journal} {Nat. Nano.}\ }\textbf {\bibinfo {volume}
  {7}},\ \bibinfo {pages} {490} (\bibinfo {year} {2012})}\BibitemShut {NoStop}%
\bibitem [{\citenamefont {Xiao}\ \emph {et~al.}(2012)\citenamefont {Xiao},
  \citenamefont {Liu}, \citenamefont {Feng}, \citenamefont {Xu},\ and\
  \citenamefont {Yao}}]{l108196802}%
  \BibitemOpen
  \bibfield  {author} {\bibinfo {author} {\bibfnamefont {D.}~\bibnamefont
  {Xiao}}, \bibinfo {author} {\bibfnamefont {G.~B.}\ \bibnamefont {Liu}},
  \bibinfo {author} {\bibfnamefont {W.}~\bibnamefont {Feng}}, \bibinfo {author}
  {\bibfnamefont {X.}~\bibnamefont {Xu}}, \ and\ \bibinfo {author}
  {\bibfnamefont {W.}~\bibnamefont {Yao}},\ }\href@noop {} {\bibfield
  {journal} {\bibinfo  {journal} {Phys. Rev. Lett.}\ }\textbf {\bibinfo
  {volume} {108}},\ \bibinfo {pages} {196802} (\bibinfo {year}
  {2012})}\BibitemShut {NoStop}%
\bibitem [{\citenamefont {Sallen}\ \emph {et~al.}(2012)\citenamefont {Sallen},
  \citenamefont {Bouet}, \citenamefont {Marie}, \citenamefont {Wang},
  \citenamefont {Zhu}, \citenamefont {Han}, \citenamefont {Lu}, \citenamefont
  {Tan}, \citenamefont {Amand}, \citenamefont {Liu},\ and\ \citenamefont
  {Urbaszek}}]{b86081301}%
  \BibitemOpen
  \bibfield  {author} {\bibinfo {author} {\bibfnamefont {G.}~\bibnamefont
  {Sallen}}, \bibinfo {author} {\bibfnamefont {L.}~\bibnamefont {Bouet}},
  \bibinfo {author} {\bibfnamefont {X.}~\bibnamefont {Marie}}, \bibinfo
  {author} {\bibfnamefont {G.}~\bibnamefont {Wang}}, \bibinfo {author}
  {\bibfnamefont {C.~R.}\ \bibnamefont {Zhu}}, \bibinfo {author} {\bibfnamefont
  {W.~P.}\ \bibnamefont {Han}}, \bibinfo {author} {\bibfnamefont
  {Y.}~\bibnamefont {Lu}}, \bibinfo {author} {\bibfnamefont {P.~H.}\
  \bibnamefont {Tan}}, \bibinfo {author} {\bibfnamefont {T.}~\bibnamefont
  {Amand}}, \bibinfo {author} {\bibfnamefont {B.~L.}\ \bibnamefont {Liu}}, \
  and\ \bibinfo {author} {\bibfnamefont {B.}~\bibnamefont {Urbaszek}},\
  }\href@noop {} {\bibfield  {journal} {\bibinfo  {journal} {Phys. Rev. B}\
  }\textbf {\bibinfo {volume} {86}},\ \bibinfo {pages} {081301} (\bibinfo
  {year} {2012})}\BibitemShut {NoStop}%
\bibitem [{\citenamefont {Benameur}\ \emph {et~al.}(2011)\citenamefont
  {Benameur}, \citenamefont {Radisavljevic}, \citenamefont
  {H$\mathrm{\acute{e}}$ron}, \citenamefont {Sahoo}, \citenamefont {Berger},\
  and\ \citenamefont {Kis}}]{Nanotechnology22125706}%
  \BibitemOpen
  \bibfield  {author} {\bibinfo {author} {\bibfnamefont {M.~M.}\ \bibnamefont
  {Benameur}}, \bibinfo {author} {\bibfnamefont {B.}~\bibnamefont
  {Radisavljevic}}, \bibinfo {author} {\bibfnamefont {J.~S.}\ \bibnamefont
  {H$\mathrm{\acute{e}}$ron}}, \bibinfo {author} {\bibfnamefont
  {S.}~\bibnamefont {Sahoo}}, \bibinfo {author} {\bibfnamefont
  {H.}~\bibnamefont {Berger}}, \ and\ \bibinfo {author} {\bibfnamefont
  {A.}~\bibnamefont {Kis}},\ }\href@noop {} {\bibfield  {journal} {\bibinfo
  {journal} {Nanotechnology}\ }\textbf {\bibinfo {volume} {22}},\ \bibinfo
  {pages} {125706} (\bibinfo {year} {2011})}\BibitemShut {NoStop}%
\bibitem [{\citenamefont {Li}\ \emph {et~al.}(2012)\citenamefont {Li},
  \citenamefont {Lu}, \citenamefont {Yin}, \citenamefont {He}, \citenamefont
  {Li}, \citenamefont {Zhang},\ and\ \citenamefont {Zhang}}]{small8682}%
  \BibitemOpen
  \bibfield  {author} {\bibinfo {author} {\bibfnamefont {H.}~\bibnamefont
  {Li}}, \bibinfo {author} {\bibfnamefont {G.}~\bibnamefont {Lu}}, \bibinfo
  {author} {\bibfnamefont {Z.~Y.}\ \bibnamefont {Yin}}, \bibinfo {author}
  {\bibfnamefont {Q.~Y.}\ \bibnamefont {He}}, \bibinfo {author} {\bibfnamefont
  {H.}~\bibnamefont {Li}}, \bibinfo {author} {\bibfnamefont {Q.}~\bibnamefont
  {Zhang}}, \ and\ \bibinfo {author} {\bibfnamefont {H.}~\bibnamefont
  {Zhang}},\ }\href@noop {} {\bibfield  {journal} {\bibinfo  {journal} {Small}\
  }\textbf {\bibinfo {volume} {8}},\ \bibinfo {pages} {682} (\bibinfo {year}
  {2012})}\BibitemShut {NoStop}%
\bibitem [{\citenamefont {Castellanos-Gomez}\ \emph {et~al.}(2010)\citenamefont
  {Castellanos-Gomez}, \citenamefont {Agra\"it},\ and\ \citenamefont
  {Rubio-Bollinger}}]{apl96213116}%
  \BibitemOpen
  \bibfield  {author} {\bibinfo {author} {\bibfnamefont {A.}~\bibnamefont
  {Castellanos-Gomez}}, \bibinfo {author} {\bibfnamefont {N.}~\bibnamefont
  {Agra\"it}}, \ and\ \bibinfo {author} {\bibfnamefont {G.}~\bibnamefont
  {Rubio-Bollinger}},\ }\href@noop {} {\bibfield  {journal} {\bibinfo
  {journal} {Appl. Phys. Lett.}\ }\textbf {\bibinfo {volume} {96}},\ \bibinfo
  {pages} {213116} (\bibinfo {year} {2010})}\BibitemShut {NoStop}%
\bibitem [{\citenamefont {Lee}\ \emph {et~al.}(2010)\citenamefont {Lee},
  \citenamefont {Yan}, \citenamefont {Brus}, \citenamefont {Heinz},
  \citenamefont {Hone},\ and\ \citenamefont {Ryu}}]{acsnano42695}%
  \BibitemOpen
  \bibfield  {author} {\bibinfo {author} {\bibfnamefont {C.}~\bibnamefont
  {Lee}}, \bibinfo {author} {\bibfnamefont {H.}~\bibnamefont {Yan}}, \bibinfo
  {author} {\bibfnamefont {L.~E.}\ \bibnamefont {Brus}}, \bibinfo {author}
  {\bibfnamefont {T.~F.}\ \bibnamefont {Heinz}}, \bibinfo {author}
  {\bibfnamefont {J.}~\bibnamefont {Hone}}, \ and\ \bibinfo {author}
  {\bibfnamefont {S.}~\bibnamefont {Ryu}},\ }\href@noop {} {\bibfield
  {journal} {\bibinfo  {journal} {{ACS Nano}}\ }\textbf {\bibinfo {volume}
  {4}},\ \bibinfo {pages} {2695} (\bibinfo {year} {2010})}\BibitemShut
  {NoStop}%
\bibitem [{\citenamefont {Molina-Sanchez}\ and\ \citenamefont
  {Wirtz}(2011)}]{b84155413}%
  \BibitemOpen
  \bibfield  {author} {\bibinfo {author} {\bibfnamefont {A.}~\bibnamefont
  {Molina-Sanchez}}\ and\ \bibinfo {author} {\bibfnamefont {L.}~\bibnamefont
  {Wirtz}},\ }\href@noop {} {\bibfield  {journal} {\bibinfo  {journal} {Phys.
  Rev. B}\ }\textbf {\bibinfo {volume} {84}},\ \bibinfo {pages} {155413}
  (\bibinfo {year} {2011})}\BibitemShut {NoStop}%
\bibitem [{\citenamefont {{Zeng}}\ \emph {et~al.}(2012)\citenamefont {{Zeng}},
  \citenamefont {{Liu}}, \citenamefont {{Dai}}, \citenamefont {{Yan}},
  \citenamefont {{Zhu}}, \citenamefont {{He}}, \citenamefont {{Xie}},
  \citenamefont {{Xu}}, \citenamefont {{Chen}}, \citenamefont {{Yao}},\ and\
  \citenamefont {{Cui}}}]{arXiv12085864}%
  \BibitemOpen
  \bibfield  {author} {\bibinfo {author} {\bibfnamefont {H.}~\bibnamefont
  {{Zeng}}}, \bibinfo {author} {\bibfnamefont {G.-B.}\ \bibnamefont {{Liu}}},
  \bibinfo {author} {\bibfnamefont {J.}~\bibnamefont {{Dai}}}, \bibinfo
  {author} {\bibfnamefont {Y.}~\bibnamefont {{Yan}}}, \bibinfo {author}
  {\bibfnamefont {B.}~\bibnamefont {{Zhu}}}, \bibinfo {author} {\bibfnamefont
  {R.}~\bibnamefont {{He}}}, \bibinfo {author} {\bibfnamefont {L.}~\bibnamefont
  {{Xie}}}, \bibinfo {author} {\bibfnamefont {S.}~\bibnamefont {{Xu}}},
  \bibinfo {author} {\bibfnamefont {X.}~\bibnamefont {{Chen}}}, \bibinfo
  {author} {\bibfnamefont {W.}~\bibnamefont {{Yao}}}, \ and\ \bibinfo {author}
  {\bibfnamefont {X.}~\bibnamefont {{Cui}}},\ }\href@noop {} {\bibfield
  {journal} {\bibinfo  {journal} {ArXiv e-prints}\ } (\bibinfo {year}
  {2012})},\ \Eprint {http://arxiv.org/abs/1208.5864} {arXiv:1208.5864}
  \BibitemShut {NoStop}%
\bibitem [{\citenamefont {Novoselov}\ \emph {et~al.}(2005)\citenamefont
  {Novoselov}, \citenamefont {Jiang}, \citenamefont {Schedin}, \citenamefont
  {Booth}, \citenamefont {Khotkevich}, \citenamefont {Morozov},\ and\
  \citenamefont {Geim}}]{pnas10210451}%
  \BibitemOpen
  \bibfield  {author} {\bibinfo {author} {\bibfnamefont {K.~S.}\ \bibnamefont
  {Novoselov}}, \bibinfo {author} {\bibfnamefont {D.}~\bibnamefont {Jiang}},
  \bibinfo {author} {\bibfnamefont {F.}~\bibnamefont {Schedin}}, \bibinfo
  {author} {\bibfnamefont {T.~J.}\ \bibnamefont {Booth}}, \bibinfo {author}
  {\bibfnamefont {V.~V.}\ \bibnamefont {Khotkevich}}, \bibinfo {author}
  {\bibfnamefont {S.~V.}\ \bibnamefont {Morozov}}, \ and\ \bibinfo {author}
  {\bibfnamefont {A.~K.}\ \bibnamefont {Geim}},\ }\href@noop {} {\bibfield
  {journal} {\bibinfo  {journal} {Proc. Natl. Acad. Sci. U.S.A.}\ }\textbf
  {\bibinfo {volume} {102}},\ \bibinfo {pages} {10451} (\bibinfo {year}
  {2005})}\BibitemShut {NoStop}%
\bibitem [{\citenamefont {Eda}\ \emph {et~al.}(2011)\citenamefont {Eda},
  \citenamefont {Yamaguchi}, \citenamefont {Voiry}, \citenamefont {Fujita},
  \citenamefont {Chen},\ and\ \citenamefont {Chhowalla}}]{nl115111}%
  \BibitemOpen
  \bibfield  {author} {\bibinfo {author} {\bibfnamefont {G.}~\bibnamefont
  {Eda}}, \bibinfo {author} {\bibfnamefont {H.}~\bibnamefont {Yamaguchi}},
  \bibinfo {author} {\bibfnamefont {D.}~\bibnamefont {Voiry}}, \bibinfo
  {author} {\bibfnamefont {T.}~\bibnamefont {Fujita}}, \bibinfo {author}
  {\bibfnamefont {M.}~\bibnamefont {Chen}}, \ and\ \bibinfo {author}
  {\bibfnamefont {M.}~\bibnamefont {Chhowalla}},\ }\href@noop {} {\bibfield
  {journal} {\bibinfo  {journal} {Nano Lett.}\ }\textbf {\bibinfo {volume}
  {11}},\ \bibinfo {pages} {5111} (\bibinfo {year} {2011})}\BibitemShut
  {NoStop}%
\bibitem [{\citenamefont {Zeng}\ \emph {et~al.}(2011)\citenamefont {Zeng},
  \citenamefont {Yin}, \citenamefont {Huang}, \citenamefont {Li}, \citenamefont
  {He}, \citenamefont {Lu}, \citenamefont {Boey},\ and\ \citenamefont
  {Zhang}}]{acie5011093}%
  \BibitemOpen
  \bibfield  {author} {\bibinfo {author} {\bibfnamefont {Z.}~\bibnamefont
  {Zeng}}, \bibinfo {author} {\bibfnamefont {Z.}~\bibnamefont {Yin}}, \bibinfo
  {author} {\bibfnamefont {X.}~\bibnamefont {Huang}}, \bibinfo {author}
  {\bibfnamefont {H.}~\bibnamefont {Li}}, \bibinfo {author} {\bibfnamefont
  {Q.}~\bibnamefont {He}}, \bibinfo {author} {\bibfnamefont {G.}~\bibnamefont
  {Lu}}, \bibinfo {author} {\bibfnamefont {F.}~\bibnamefont {Boey}}, \ and\
  \bibinfo {author} {\bibfnamefont {H.}~\bibnamefont {Zhang}},\ }\href@noop {}
  {\bibfield  {journal} {\bibinfo  {journal} {Angew. Chem. Int. Ed.}\ }\textbf
  {\bibinfo {volume} {50}},\ \bibinfo {pages} {11093} (\bibinfo {year}
  {2011})}\BibitemShut {NoStop}%
\bibitem [{\citenamefont {Dines}(1975)}]{mrs10287}%
  \BibitemOpen
  \bibfield  {author} {\bibinfo {author} {\bibfnamefont {M.~B.}\ \bibnamefont
  {Dines}},\ }\href@noop {} {\bibfield  {journal} {\bibinfo  {journal} {Mater.
  Res. Bul.}\ }\textbf {\bibinfo {volume} {10}},\ \bibinfo {pages} {287}
  (\bibinfo {year} {1975})}\BibitemShut {NoStop}%
\bibitem [{\citenamefont {Frey}\ \emph {et~al.}(2003)\citenamefont {Frey},
  \citenamefont {Reynolds}, \citenamefont {Friend}, \citenamefont {Cohen},\
  and\ \citenamefont {Feldman}}]{jacs1255998}%
  \BibitemOpen
  \bibfield  {author} {\bibinfo {author} {\bibfnamefont {G.~L.}\ \bibnamefont
  {Frey}}, \bibinfo {author} {\bibfnamefont {K.~J.}\ \bibnamefont {Reynolds}},
  \bibinfo {author} {\bibfnamefont {R.~H.}\ \bibnamefont {Friend}}, \bibinfo
  {author} {\bibfnamefont {H.}~\bibnamefont {Cohen}}, \ and\ \bibinfo {author}
  {\bibfnamefont {Y.}~\bibnamefont {Feldman}},\ }\href@noop {} {\bibfield
  {journal} {\bibinfo  {journal} {J. Am. Chem. Soc.}\ }\textbf {\bibinfo
  {volume} {125}},\ \bibinfo {pages} {5998} (\bibinfo {year}
  {2003})}\BibitemShut {NoStop}%
\bibitem [{\citenamefont {Zeng}\ \emph {et~al.}(2012)\citenamefont {Zeng},
  \citenamefont {Sun}, \citenamefont {Zhu}, \citenamefont {Huang},
  \citenamefont {Yin}, \citenamefont {Lu}, \citenamefont {Fan}, \citenamefont
  {Yan}, \citenamefont {Hng},\ and\ \citenamefont {Zhang}}]{acie519052}%
  \BibitemOpen
  \bibfield  {author} {\bibinfo {author} {\bibfnamefont {Z.}~\bibnamefont
  {Zeng}}, \bibinfo {author} {\bibfnamefont {T.}~\bibnamefont {Sun}}, \bibinfo
  {author} {\bibfnamefont {J.}~\bibnamefont {Zhu}}, \bibinfo {author}
  {\bibfnamefont {X.}~\bibnamefont {Huang}}, \bibinfo {author} {\bibfnamefont
  {Z.}~\bibnamefont {Yin}}, \bibinfo {author} {\bibfnamefont {G.}~\bibnamefont
  {Lu}}, \bibinfo {author} {\bibfnamefont {Z.}~\bibnamefont {Fan}}, \bibinfo
  {author} {\bibfnamefont {Q.}~\bibnamefont {Yan}}, \bibinfo {author}
  {\bibfnamefont {H.~H.}\ \bibnamefont {Hng}}, \ and\ \bibinfo {author}
  {\bibfnamefont {H.}~\bibnamefont {Zhang}},\ }\href@noop {} {\bibfield
  {journal} {\bibinfo  {journal} {Angew. Chem. Int. Ed.}\ }\textbf {\bibinfo
  {volume} {51}},\ \bibinfo {pages} {9052} (\bibinfo {year}
  {2012})}\BibitemShut {NoStop}%
\bibitem [{\citenamefont {Coleman}\ \emph {et~al.}(2011)\citenamefont
  {Coleman}, \citenamefont {Lotya}, \citenamefont {O'neill}, \citenamefont
  {Bergin}, \citenamefont {King}, \citenamefont {Khan}, \citenamefont {Young},
  \citenamefont {Gaucher}, \citenamefont {De}, \citenamefont {Smith},
  \citenamefont {Shvets}, \citenamefont {Arora}, \citenamefont {Stanton},
  \citenamefont {Kim}, \citenamefont {Lee}, \citenamefont {Kim}, \citenamefont
  {Duesberg}, \citenamefont {Hallam}, \citenamefont {Boland}, \citenamefont
  {Wang}, \citenamefont {Donegan}, \citenamefont {Grunlan}, \citenamefont
  {Moriarty}, \citenamefont {Shmeliov}, \citenamefont {Nicholls}, \citenamefont
  {Perkins}, \citenamefont {Grieveson}, \citenamefont {Theuwissen},
  \citenamefont {{McComb}}, \citenamefont {Nellist},\ and\ \citenamefont
  {Nicolosi}}]{s331568}%
  \BibitemOpen
  \bibfield  {author} {\bibinfo {author} {\bibfnamefont {J.~N.}\ \bibnamefont
  {Coleman}}, \bibinfo {author} {\bibfnamefont {M.}~\bibnamefont {Lotya}},
  \bibinfo {author} {\bibfnamefont {A.}~\bibnamefont {O'neill}}, \bibinfo
  {author} {\bibfnamefont {S.~D.}\ \bibnamefont {Bergin}}, \bibinfo {author}
  {\bibfnamefont {P.~J.}\ \bibnamefont {King}}, \bibinfo {author}
  {\bibfnamefont {U.}~\bibnamefont {Khan}}, \bibinfo {author} {\bibfnamefont
  {K.}~\bibnamefont {Young}}, \bibinfo {author} {\bibfnamefont
  {A.}~\bibnamefont {Gaucher}}, \bibinfo {author} {\bibfnamefont
  {S.}~\bibnamefont {De}}, \bibinfo {author} {\bibfnamefont {R.~J.}\
  \bibnamefont {Smith}}, \bibinfo {author} {\bibfnamefont {I.~V.}\ \bibnamefont
  {Shvets}}, \bibinfo {author} {\bibfnamefont {S.~K.}\ \bibnamefont {Arora}},
  \bibinfo {author} {\bibfnamefont {G.}~\bibnamefont {Stanton}}, \bibinfo
  {author} {\bibfnamefont {H.-Y.}\ \bibnamefont {Kim}}, \bibinfo {author}
  {\bibfnamefont {K.}~\bibnamefont {Lee}}, \bibinfo {author} {\bibfnamefont
  {G.~T.}\ \bibnamefont {Kim}}, \bibinfo {author} {\bibfnamefont {G.~S.}\
  \bibnamefont {Duesberg}}, \bibinfo {author} {\bibfnamefont {T.}~\bibnamefont
  {Hallam}}, \bibinfo {author} {\bibfnamefont {J.~J.}\ \bibnamefont {Boland}},
  \bibinfo {author} {\bibfnamefont {J.~J.}\ \bibnamefont {Wang}}, \bibinfo
  {author} {\bibfnamefont {J.~F.}\ \bibnamefont {Donegan}}, \bibinfo {author}
  {\bibfnamefont {J.~C.}\ \bibnamefont {Grunlan}}, \bibinfo {author}
  {\bibfnamefont {G.}~\bibnamefont {Moriarty}}, \bibinfo {author}
  {\bibfnamefont {A.}~\bibnamefont {Shmeliov}}, \bibinfo {author}
  {\bibfnamefont {R.~J.}\ \bibnamefont {Nicholls}}, \bibinfo {author}
  {\bibfnamefont {J.~M.}\ \bibnamefont {Perkins}}, \bibinfo {author}
  {\bibfnamefont {E.~M.}\ \bibnamefont {Grieveson}}, \bibinfo {author}
  {\bibfnamefont {K.}~\bibnamefont {Theuwissen}}, \bibinfo {author}
  {\bibfnamefont {D.~W.}\ \bibnamefont {{McComb}}}, \bibinfo {author}
  {\bibfnamefont {P.~D.}\ \bibnamefont {Nellist}}, \ and\ \bibinfo {author}
  {\bibfnamefont {V.}~\bibnamefont {Nicolosi}},\ }\href@noop {} {\bibfield
  {journal} {\bibinfo  {journal} {Science}\ }\textbf {\bibinfo {volume}
  {331}},\ \bibinfo {pages} {568} (\bibinfo {year} {2011})}\BibitemShut
  {NoStop}%
\bibitem [{\citenamefont {Smith}\ \emph {et~al.}(2011)\citenamefont {Smith},
  \citenamefont {King}, \citenamefont {Lotya}, \citenamefont {Wirtz},
  \citenamefont {Khan}, \citenamefont {De}, \citenamefont {O'neill},
  \citenamefont {Duesberg}, \citenamefont {Grunlan}, \citenamefont {Moriarty},
  \citenamefont {Chen}, \citenamefont {Wang}, \citenamefont {Minett},
  \citenamefont {Nicolosi},\ and\ \citenamefont {Coleman}}]{am233944}%
  \BibitemOpen
  \bibfield  {author} {\bibinfo {author} {\bibfnamefont {R.~J.}\ \bibnamefont
  {Smith}}, \bibinfo {author} {\bibfnamefont {P.~J.}\ \bibnamefont {King}},
  \bibinfo {author} {\bibfnamefont {M.}~\bibnamefont {Lotya}}, \bibinfo
  {author} {\bibfnamefont {C.}~\bibnamefont {Wirtz}}, \bibinfo {author}
  {\bibfnamefont {U.}~\bibnamefont {Khan}}, \bibinfo {author} {\bibfnamefont
  {S.}~\bibnamefont {De}}, \bibinfo {author} {\bibfnamefont {A.}~\bibnamefont
  {O'neill}}, \bibinfo {author} {\bibfnamefont {G.~S.}\ \bibnamefont
  {Duesberg}}, \bibinfo {author} {\bibfnamefont {J.~C.}\ \bibnamefont
  {Grunlan}}, \bibinfo {author} {\bibfnamefont {G.}~\bibnamefont {Moriarty}},
  \bibinfo {author} {\bibfnamefont {J.}~\bibnamefont {Chen}}, \bibinfo {author}
  {\bibfnamefont {J.}~\bibnamefont {Wang}}, \bibinfo {author} {\bibfnamefont
  {A.~I.}\ \bibnamefont {Minett}}, \bibinfo {author} {\bibfnamefont
  {V.}~\bibnamefont {Nicolosi}}, \ and\ \bibinfo {author} {\bibfnamefont
  {J.~N.}\ \bibnamefont {Coleman}},\ }\href@noop {} {\bibfield  {journal}
  {\bibinfo  {journal} {Adv. Mater.}\ }\textbf {\bibinfo {volume} {23}},\
  \bibinfo {pages} {3944} (\bibinfo {year} {2011})}\BibitemShut {NoStop}%
\bibitem [{\citenamefont {May}\ \emph {et~al.}(2012)\citenamefont {May},
  \citenamefont {Khan}, \citenamefont {Hughes},\ and\ \citenamefont
  {Coleman}}]{jpcc11611393}%
  \BibitemOpen
  \bibfield  {author} {\bibinfo {author} {\bibfnamefont {P.}~\bibnamefont
  {May}}, \bibinfo {author} {\bibfnamefont {U.}~\bibnamefont {Khan}}, \bibinfo
  {author} {\bibfnamefont {J.~M.}\ \bibnamefont {Hughes}}, \ and\ \bibinfo
  {author} {\bibfnamefont {J.~N.}\ \bibnamefont {Coleman}},\ }\href@noop {}
  {\bibfield  {journal} {\bibinfo  {journal} {J. Phys. Chem. C}\ }\textbf
  {\bibinfo {volume} {116}},\ \bibinfo {pages} {11393} (\bibinfo {year}
  {2012})}\BibitemShut {NoStop}%
\bibitem [{\citenamefont {Zhou}\ \emph {et~al.}(2011)\citenamefont {Zhou},
  \citenamefont {Mao}, \citenamefont {Wang}, \citenamefont {Peng},\ and\
  \citenamefont {Zhang}}]{acie5010839}%
  \BibitemOpen
  \bibfield  {author} {\bibinfo {author} {\bibfnamefont {K.-G.}\ \bibnamefont
  {Zhou}}, \bibinfo {author} {\bibfnamefont {N.-N.}\ \bibnamefont {Mao}},
  \bibinfo {author} {\bibfnamefont {H.-X.}\ \bibnamefont {Wang}}, \bibinfo
  {author} {\bibfnamefont {Y.}~\bibnamefont {Peng}}, \ and\ \bibinfo {author}
  {\bibfnamefont {H.-L.}\ \bibnamefont {Zhang}},\ }\href@noop {} {\bibfield
  {journal} {\bibinfo  {journal} {Angew. Chem. Int. Ed.}\ }\textbf {\bibinfo
  {volume} {50}},\ \bibinfo {pages} {10839} (\bibinfo {year}
  {2011})}\BibitemShut {NoStop}%
\bibitem [{\citenamefont {Cunningham}\ \emph {et~al.}(2012)\citenamefont
  {Cunningham}, \citenamefont {Lotya}, \citenamefont {Cucinotta}, \citenamefont
  {Sanvito}, \citenamefont {Bergin}, \citenamefont {Menzel}, \citenamefont
  {Shaffer},\ and\ \citenamefont {Coleman}}]{acsnano63468}%
  \BibitemOpen
  \bibfield  {author} {\bibinfo {author} {\bibfnamefont {G.}~\bibnamefont
  {Cunningham}}, \bibinfo {author} {\bibfnamefont {M.}~\bibnamefont {Lotya}},
  \bibinfo {author} {\bibfnamefont {C.~S.}\ \bibnamefont {Cucinotta}}, \bibinfo
  {author} {\bibfnamefont {S.}~\bibnamefont {Sanvito}}, \bibinfo {author}
  {\bibfnamefont {S.~D.}\ \bibnamefont {Bergin}}, \bibinfo {author}
  {\bibfnamefont {R.}~\bibnamefont {Menzel}}, \bibinfo {author} {\bibfnamefont
  {M.~S.~P.}\ \bibnamefont {Shaffer}}, \ and\ \bibinfo {author} {\bibfnamefont
  {J.~N.}\ \bibnamefont {Coleman}},\ }\href@noop {} {\bibfield  {journal}
  {\bibinfo  {journal} {ACS Nano}\ }\textbf {\bibinfo {volume} {6}},\ \bibinfo
  {pages} {3468} (\bibinfo {year} {2012})}\BibitemShut {NoStop}%
\bibitem [{\citenamefont {Altavilla}\ \emph {et~al.}(2011)\citenamefont
  {Altavilla}, \citenamefont {Sarno},\ and\ \citenamefont
  {Ciambelli}}]{cm233879}%
  \BibitemOpen
  \bibfield  {author} {\bibinfo {author} {\bibfnamefont {C.}~\bibnamefont
  {Altavilla}}, \bibinfo {author} {\bibfnamefont {M.}~\bibnamefont {Sarno}}, \
  and\ \bibinfo {author} {\bibfnamefont {P.}~\bibnamefont {Ciambelli}},\
  }\href@noop {} {\bibfield  {journal} {\bibinfo  {journal} {Chem. Mate.}\
  }\textbf {\bibinfo {volume} {23}},\ \bibinfo {pages} {3879} (\bibinfo {year}
  {2011})}\BibitemShut {NoStop}%
\bibitem [{\citenamefont {O'neill}\ \emph {et~al.}(2012)\citenamefont
  {O'neill}, \citenamefont {Khan},\ and\ \citenamefont {Coleman}}]{cm242414}%
  \BibitemOpen
  \bibfield  {author} {\bibinfo {author} {\bibfnamefont {A.}~\bibnamefont
  {O'neill}}, \bibinfo {author} {\bibfnamefont {U.}~\bibnamefont {Khan}}, \
  and\ \bibinfo {author} {\bibfnamefont {J.~N.}\ \bibnamefont {Coleman}},\
  }\href@noop {} {\bibfield  {journal} {\bibinfo  {journal} {Chem. Mater.}\
  }\textbf {\bibinfo {volume} {24}},\ \bibinfo {pages} {2414} (\bibinfo {year}
  {2012})}\BibitemShut {NoStop}%
\bibitem [{\citenamefont {Peng}\ \emph
  {et~al.}(2001{\natexlab{a}})\citenamefont {Peng}, \citenamefont {Meng},
  \citenamefont {Zhong}, \citenamefont {Lu}, \citenamefont {Yu}, \citenamefont
  {Yang},\ and\ \citenamefont {Qian}}]{jssc159170}%
  \BibitemOpen
  \bibfield  {author} {\bibinfo {author} {\bibfnamefont {Y.~Y.}\ \bibnamefont
  {Peng}}, \bibinfo {author} {\bibfnamefont {Z.~Y.}\ \bibnamefont {Meng}},
  \bibinfo {author} {\bibfnamefont {C.}~\bibnamefont {Zhong}}, \bibinfo
  {author} {\bibfnamefont {J.}~\bibnamefont {Lu}}, \bibinfo {author}
  {\bibfnamefont {W.~C.}\ \bibnamefont {Yu}}, \bibinfo {author} {\bibfnamefont
  {Z.~P.}\ \bibnamefont {Yang}}, \ and\ \bibinfo {author} {\bibfnamefont
  {Y.~T.}\ \bibnamefont {Qian}},\ }\href@noop {} {\bibfield  {journal}
  {\bibinfo  {journal} {J. Solid. State. Chem.}\ }\textbf {\bibinfo {volume}
  {159}},\ \bibinfo {pages} {170} (\bibinfo {year}
  {2001}{\natexlab{a}})}\BibitemShut {NoStop}%
\bibitem [{\citenamefont {Peng}\ \emph
  {et~al.}(2001{\natexlab{b}})\citenamefont {Peng}, \citenamefont {Meng},
  \citenamefont {Zhong}, \citenamefont {Lu}, \citenamefont {Yu}, \citenamefont
  {Jia},\ and\ \citenamefont {Qian}}]{cl30772}%
  \BibitemOpen
  \bibfield  {author} {\bibinfo {author} {\bibfnamefont {Y.~Y.}\ \bibnamefont
  {Peng}}, \bibinfo {author} {\bibfnamefont {Z.~Y.}\ \bibnamefont {Meng}},
  \bibinfo {author} {\bibfnamefont {C.}~\bibnamefont {Zhong}}, \bibinfo
  {author} {\bibfnamefont {J.}~\bibnamefont {Lu}}, \bibinfo {author}
  {\bibfnamefont {W.~C.}\ \bibnamefont {Yu}}, \bibinfo {author} {\bibfnamefont
  {Y.~B.}\ \bibnamefont {Jia}}, \ and\ \bibinfo {author} {\bibfnamefont
  {Y.~T.}\ \bibnamefont {Qian}},\ }\href@noop {} {\bibfield  {journal}
  {\bibinfo  {journal} {Chem. Lett.}\ }\textbf {\bibinfo {volume} {30}},\
  \bibinfo {pages} {772} (\bibinfo {year} {2001}{\natexlab{b}})}\BibitemShut
  {NoStop}%
\bibitem [{\citenamefont {Lee}\ \emph {et~al.}(2012{\natexlab{b}})\citenamefont
  {Lee}, \citenamefont {Zhang}, \citenamefont {Zhang}, \citenamefont {Chang},
  \citenamefont {Lin}, \citenamefont {Chang}, \citenamefont {Yu}, \citenamefont
  {Wang}, \citenamefont {Chang}, \citenamefont {Li},\ and\ \citenamefont
  {Lin}}]{am242320}%
  \BibitemOpen
  \bibfield  {author} {\bibinfo {author} {\bibfnamefont {Y.~H.}\ \bibnamefont
  {Lee}}, \bibinfo {author} {\bibfnamefont {X.~Q.}\ \bibnamefont {Zhang}},
  \bibinfo {author} {\bibfnamefont {W.}~\bibnamefont {Zhang}}, \bibinfo
  {author} {\bibfnamefont {M.~T.}\ \bibnamefont {Chang}}, \bibinfo {author}
  {\bibfnamefont {C.~T.}\ \bibnamefont {Lin}}, \bibinfo {author} {\bibfnamefont
  {K.~D.}\ \bibnamefont {Chang}}, \bibinfo {author} {\bibfnamefont {Y.~C.}\
  \bibnamefont {Yu}}, \bibinfo {author} {\bibfnamefont {J.~T.}\ \bibnamefont
  {Wang}}, \bibinfo {author} {\bibfnamefont {C.~S.}\ \bibnamefont {Chang}},
  \bibinfo {author} {\bibfnamefont {L.~J.}\ \bibnamefont {Li}}, \ and\ \bibinfo
  {author} {\bibfnamefont {T.~W.}\ \bibnamefont {Lin}},\ }\href@noop {}
  {\bibfield  {journal} {\bibinfo  {journal} {Adv. Mater.}\ }\textbf {\bibinfo
  {volume} {24}},\ \bibinfo {pages} {2320} (\bibinfo {year}
  {2012}{\natexlab{b}})}\BibitemShut {NoStop}%
\bibitem [{\citenamefont {Zhan}\ \emph {et~al.}(2012)\citenamefont {Zhan},
  \citenamefont {Liu}, \citenamefont {Najmaei}, \citenamefont {Ajayan},\ and\
  \citenamefont {Lou}}]{small8966}%
  \BibitemOpen
  \bibfield  {author} {\bibinfo {author} {\bibfnamefont {Y.~J.}\ \bibnamefont
  {Zhan}}, \bibinfo {author} {\bibfnamefont {Z.}~\bibnamefont {Liu}}, \bibinfo
  {author} {\bibfnamefont {S.}~\bibnamefont {Najmaei}}, \bibinfo {author}
  {\bibfnamefont {P.~M.}\ \bibnamefont {Ajayan}}, \ and\ \bibinfo {author}
  {\bibfnamefont {J.}~\bibnamefont {Lou}},\ }\href@noop {} {\bibfield
  {journal} {\bibinfo  {journal} {Small}\ }\textbf {\bibinfo {volume} {8}},\
  \bibinfo {pages} {966} (\bibinfo {year} {2012})}\BibitemShut {NoStop}%
\bibitem [{\citenamefont {Liu}\ \emph {et~al.}(2012)\citenamefont {Liu},
  \citenamefont {Zhang}, \citenamefont {Lee}, \citenamefont {Lin},
  \citenamefont {Chang}, \citenamefont {Su}, \citenamefont {Chang},
  \citenamefont {Li}, \citenamefont {Shi}, \citenamefont {Zhang}, \citenamefont
  {Lai},\ and\ \citenamefont {Li}}]{nl121538}%
  \BibitemOpen
  \bibfield  {author} {\bibinfo {author} {\bibfnamefont {K.~K.}\ \bibnamefont
  {Liu}}, \bibinfo {author} {\bibfnamefont {W.}~\bibnamefont {Zhang}}, \bibinfo
  {author} {\bibfnamefont {Y.~H.}\ \bibnamefont {Lee}}, \bibinfo {author}
  {\bibfnamefont {Y.~C.}\ \bibnamefont {Lin}}, \bibinfo {author} {\bibfnamefont
  {M.~T.}\ \bibnamefont {Chang}}, \bibinfo {author} {\bibfnamefont {C.~Y.}\
  \bibnamefont {Su}}, \bibinfo {author} {\bibfnamefont {C.~S.}\ \bibnamefont
  {Chang}}, \bibinfo {author} {\bibfnamefont {H.}~\bibnamefont {Li}}, \bibinfo
  {author} {\bibfnamefont {Y.}~\bibnamefont {Shi}}, \bibinfo {author}
  {\bibfnamefont {H.}~\bibnamefont {Zhang}}, \bibinfo {author} {\bibfnamefont
  {C.~S.}\ \bibnamefont {Lai}}, \ and\ \bibinfo {author} {\bibfnamefont
  {L.~J.}\ \bibnamefont {Li}},\ }\href@noop {} {\bibfield  {journal} {\bibinfo
  {journal} {Nano Lett.}\ }\textbf {\bibinfo {volume} {12}},\ \bibinfo {pages}
  {1538} (\bibinfo {year} {2012})}\BibitemShut {NoStop}%
\bibitem [{\citenamefont {Balendhran}\ \emph {et~al.}(2012)\citenamefont
  {Balendhran}, \citenamefont {Ou}, \citenamefont {Bhaskaran}, \citenamefont
  {Sriram}, \citenamefont {Ippolito}, \citenamefont {Vasic}, \citenamefont
  {Kats}, \citenamefont {Bhargava}, \citenamefont {Zhuiykov},\ and\
  \citenamefont {Kalantar-Zadeh}}]{nanoscale4461}%
  \BibitemOpen
  \bibfield  {author} {\bibinfo {author} {\bibfnamefont {S.}~\bibnamefont
  {Balendhran}}, \bibinfo {author} {\bibfnamefont {J.~Z.}\ \bibnamefont {Ou}},
  \bibinfo {author} {\bibfnamefont {M.}~\bibnamefont {Bhaskaran}}, \bibinfo
  {author} {\bibfnamefont {S.}~\bibnamefont {Sriram}}, \bibinfo {author}
  {\bibfnamefont {S.}~\bibnamefont {Ippolito}}, \bibinfo {author}
  {\bibfnamefont {Z.}~\bibnamefont {Vasic}}, \bibinfo {author} {\bibfnamefont
  {E.}~\bibnamefont {Kats}}, \bibinfo {author} {\bibfnamefont {S.}~\bibnamefont
  {Bhargava}}, \bibinfo {author} {\bibfnamefont {S.}~\bibnamefont {Zhuiykov}},
  \ and\ \bibinfo {author} {\bibfnamefont {K.}~\bibnamefont {Kalantar-Zadeh}},\
  }\href@noop {} {\bibfield  {journal} {\bibinfo  {journal} {Nanoscale}\
  }\textbf {\bibinfo {volume} {4}},\ \bibinfo {pages} {461} (\bibinfo {year}
  {2012})}\BibitemShut {NoStop}%
\bibitem [{\citenamefont {Shi}\ \emph {et~al.}(2012)\citenamefont {Shi},
  \citenamefont {Zhou}, \citenamefont {Lu}, \citenamefont {Fang}, \citenamefont
  {Lee}, \citenamefont {Hsu}, \citenamefont {Kim}, \citenamefont {Kim},
  \citenamefont {Yang}, \citenamefont {Li}, \citenamefont {Idrobo},\ and\
  \citenamefont {Kong}}]{nl122784}%
  \BibitemOpen
  \bibfield  {author} {\bibinfo {author} {\bibfnamefont {Y.~M.}\ \bibnamefont
  {Shi}}, \bibinfo {author} {\bibfnamefont {W.}~\bibnamefont {Zhou}}, \bibinfo
  {author} {\bibfnamefont {A.~Y.}\ \bibnamefont {Lu}}, \bibinfo {author}
  {\bibfnamefont {W.~J.}\ \bibnamefont {Fang}}, \bibinfo {author}
  {\bibfnamefont {Y.~H.}\ \bibnamefont {Lee}}, \bibinfo {author} {\bibfnamefont
  {A.~L.}\ \bibnamefont {Hsu}}, \bibinfo {author} {\bibfnamefont {S.~M.}\
  \bibnamefont {Kim}}, \bibinfo {author} {\bibfnamefont {K.~K.}\ \bibnamefont
  {Kim}}, \bibinfo {author} {\bibfnamefont {H.~Y.}\ \bibnamefont {Yang}},
  \bibinfo {author} {\bibfnamefont {L.~J.}\ \bibnamefont {Li}}, \bibinfo
  {author} {\bibfnamefont {J.~C.}\ \bibnamefont {Idrobo}}, \ and\ \bibinfo
  {author} {\bibfnamefont {J.}~\bibnamefont {Kong}},\ }\href@noop {} {\bibfield
   {journal} {\bibinfo  {journal} {Nano Lett.}\ }\textbf {\bibinfo {volume}
  {12}},\ \bibinfo {pages} {2784} (\bibinfo {year} {2012})}\BibitemShut
  {NoStop}%
\bibitem [{\citenamefont {{Najmaei}}\ \emph {et~al.}(2013)\citenamefont
  {{Najmaei}}, \citenamefont {{Liu}}, \citenamefont {{Zhou}}, \citenamefont
  {{Zou}}, \citenamefont {{Shi}}, \citenamefont {{Lei}}, \citenamefont
  {{Yakobson}}, \citenamefont {{Idrobo}}, \citenamefont {{Ajayan}},\ and\
  \citenamefont {{Lou}}}]{arxiv13012812}%
  \BibitemOpen
  \bibfield  {author} {\bibinfo {author} {\bibfnamefont {S.}~\bibnamefont
  {{Najmaei}}}, \bibinfo {author} {\bibfnamefont {Z.}~\bibnamefont {{Liu}}},
  \bibinfo {author} {\bibfnamefont {W.}~\bibnamefont {{Zhou}}}, \bibinfo
  {author} {\bibfnamefont {X.}~\bibnamefont {{Zou}}}, \bibinfo {author}
  {\bibfnamefont {G.}~\bibnamefont {{Shi}}}, \bibinfo {author} {\bibfnamefont
  {S.}~\bibnamefont {{Lei}}}, \bibinfo {author} {\bibfnamefont {B.~I.}\
  \bibnamefont {{Yakobson}}}, \bibinfo {author} {\bibfnamefont {J.-C.}\
  \bibnamefont {{Idrobo}}}, \bibinfo {author} {\bibfnamefont {P.~M.}\
  \bibnamefont {{Ajayan}}}, \ and\ \bibinfo {author} {\bibfnamefont
  {J.}~\bibnamefont {{Lou}}},\ }\href@noop {} {\bibfield  {journal} {\bibinfo
  {journal} {ArXiv e-prints}\ } (\bibinfo {year} {2013})},\ \Eprint
  {http://arxiv.org/abs/1301.2812} {arXiv:1301.2812} \BibitemShut {NoStop}%
\bibitem [{\citenamefont {Lauritsen}\ \emph {et~al.}(2007)\citenamefont
  {Lauritsen}, \citenamefont {Kibsgaard}, \citenamefont {Helveg}, \citenamefont
  {Topsoe}, \citenamefont {Clausen}, \citenamefont {Laegsgaard},\ and\
  \citenamefont {Besenbacher}}]{nn253}%
  \BibitemOpen
  \bibfield  {author} {\bibinfo {author} {\bibfnamefont {J.~V.}\ \bibnamefont
  {Lauritsen}}, \bibinfo {author} {\bibfnamefont {J.}~\bibnamefont
  {Kibsgaard}}, \bibinfo {author} {\bibfnamefont {S.}~\bibnamefont {Helveg}},
  \bibinfo {author} {\bibfnamefont {H.}~\bibnamefont {Topsoe}}, \bibinfo
  {author} {\bibfnamefont {B.~S.}\ \bibnamefont {Clausen}}, \bibinfo {author}
  {\bibfnamefont {E.}~\bibnamefont {Laegsgaard}}, \ and\ \bibinfo {author}
  {\bibfnamefont {F.}~\bibnamefont {Besenbacher}},\ }\href@noop {} {\bibfield
  {journal} {\bibinfo  {journal} {Nat. Nanotechnol.}\ }\textbf {\bibinfo
  {volume} {2}},\ \bibinfo {pages} {53} (\bibinfo {year} {2007})}\BibitemShut
  {NoStop}%
\bibitem [{\citenamefont {Byskov}\ \emph {et~al.}(2000)\citenamefont {Byskov},
  \citenamefont {N{\o}rskov}, \citenamefont {Clausen},\ and\ \citenamefont
  {Tops{\o}e}}]{cl6495}%
  \BibitemOpen
  \bibfield  {author} {\bibinfo {author} {\bibfnamefont {L.}~\bibnamefont
  {Byskov}}, \bibinfo {author} {\bibfnamefont {J.}~\bibnamefont {N{\o}rskov}},
  \bibinfo {author} {\bibfnamefont {B.}~\bibnamefont {Clausen}}, \ and\
  \bibinfo {author} {\bibfnamefont {H.}~\bibnamefont {Tops{\o}e}},\ }\href@noop
  {} {\bibfield  {journal} {\bibinfo  {journal} {Catalysis Letters}\ }\textbf
  {\bibinfo {volume} {64}},\ \bibinfo {pages} {95} (\bibinfo {year}
  {2000})}\BibitemShut {NoStop}%
\end{thebibliography}

%

\end{document}